\documentstyle[12pt,aasms4]{article}
\lefthead{Jun and Jones}
\righthead{Density Spike in CR-Modified-Shock}

\begin{document}

\def\etal{{\it et al.~}}
\def\eg{{\it e.g.},~}
\def\ie{{\it i.e.},~}
\def\lsim{\raise0.3ex\hbox{$<$}\kern-0.75em{\lower0.65ex\hbox{$\sim$}}}
\def\gsim{\raise0.3ex\hbox{$>$}\kern-0.75em{\lower0.65ex\hbox{$\sim$}}}

\title{The Density Spike in Cosmic-Ray-Modified Shocks: Formation,
Evolution, and Instability} 
\author{Byung-Il Jun and T.W. Jones}
\affil{Department of Astronomy, University of Minnesota \\
116 Church Street, S.E., Minneapolis, MN 55455 }

\begin{abstract}

We examine the formation and evolution of the density enhancement
(density spike) that appears downstream of strong, cosmic-ray-modified
shocks.  This feature results from temporary overcompression of the
flow by the combined cosmic-ray shock precursor/gas subshock.
Formation of 
the density spike is expected whenever shock modification by
cosmic-ray pressure increases strongly. That occurence may be
anticipated for newly generated strong shocks or for
cosmic-ray-modified shocks encountering a region of higher external
density, for example.  

The predicted mass density within the spike
increases with the shock Mach number and with shocks more
dominated  by cosmic-ray pressure.
For very strong shocks, the  total compression
compared to the upstream gas may approach ${\gamma_g +1 \over \gamma_g -1}D $
during the formation period, where $\gamma_g$ is the gas adiabatic
index and $D$ is the compression ratio through the precursor.  
As the full shock reaches equilibrium, the spike detaches, lags behind the
modified shock transition and is further compressed, so that the
density can exceed 
the limit quoted above.  We find this spike
to be linearly unstable under a modified Rayleigh-Taylor instability 
criterion at the early stage of its formation.
Our linear analysis shows that the flow is unstable when the
gradients of total pressure (gas pressure + cosmic-ray pressure) and
gas density have opposite signs.  We confirm this numerically using
two independent codes based on the two-fluid model for cosmic-ray transport.
These two-dimensional simulations show that the instability grows
impulsively at early stages and then slows down as the gradients of
total pressure and gas density decrease.
Flow within the density spike becomes disordered through
the instability. It seems likely that this can significantly increase the
local magnetic field beyond compressional effects.
Observational discovery of this unstable density spike behind shocks,
possibly through radio emission enhanced by 
the amplified magnetic fields would provide
evidence for the existence of strongly cosmic-ray modified shock structures.

\end{abstract}


\section{Introduction} % 1
  
 Energetic charged particles (Cosmic-rays or CRs) are believed to be
accelerated in 
strong shock waves through the first-order Fermi mechanism when they
become trapped there by Alfv\'en wave turbulence.
A relatively small population  of particles can achieve very high energies
by repeatedly crossing the shock and having their velocities
redistributed by scattering. 
This very efficient, diffusive acceleration mechanism was discovered
by a number of people independently in 
the late 1970s (\cite{kry77,als77,bel78,bla78}).
The process is nonlinear, since pressure feedback from accelerated CR
particles modifies the shock 
compression and adjusts the acceleration efficiency towards an
equilibrium (\eg \cite{bla80}). 
The modified shock structure includes an extended upstream pressure gradient
(``foot'' or ``precursor'') that results from upstream diffusion of
CR pressure. Within the precursor gas is adiabatically compressed.
At least during formation of the precursor there will also be a
dissipative, gas subshock. However, above some critical Mach number
diffusive shock theory suggests that
this feature can become weak or 
absent (leading to a smooth or ``C-shock'' structure) 
in CR shocks as they approach dynamical equilibrium (\eg \cite{dru86};
\cite{jon90}).  
The presence of a strong magnetic field transverse to the shock normal
tends to reduce the efficiency of CR acceleration in shocks of a
given sonic Mach number (\cite{webb86,jcn94,fran95}), so the
critical Mach number for the formation of a C-shock will generally be higher
for MHD shocks (\cite{jcn94}).

The hydromagnetic structures of CR shocks are now fairly well
studied for a variety of circumstances. Much of that effort has
focussed on steady-state conditions (\eg
\cite{dru81};~\cite{webb86};~\cite{ell90};~\cite{kan90};~\cite{jcn94}). 
On the other hand, there are important circumstances in which steady-state
CR shocks may not be applicable, such as supernova blast waves (\eg
\cite{dor84}; \cite{drumv89}; \cite{jon90}). 
One interesting feature in
time-dependent, evolving CR shocks is the appearance of a sharp density
enhancement or ``spike'' downstream of the shock transition in both
parallel and perpendicular shocks
(\cite{dor84}; \cite{dor85}; \cite{jon90}; \cite{jcn94}). This sharp density
enhancement is formed as the growth of CR pressure first enhances the
total compression through the shock and then levels off 
towards the equilibrium predicted for steady-state shocks. The maximum
density within the spike can exceed that for the equilibrium
modified shock and the initial shock by a large factor when the
shock Mach number is large.
We have discovered that this feature is Rayleigh-Taylor unstable.
That could have interesting observable consequences, if, for
example, it lead to amplification of magnetic fields within a well-defined
region behind a supernova shock.
In this paper, we present one- and two-dimensional
numerical simulations of the evolution of the density spike, demonstrating its
instability.  We point out that the instability discussed here is
distinct from other previously reported instabilities associated
with the CR shock structure itself. On the other hand, the instability
requires a large CR pressure and
occurs because of modifications to the shock. Our
simulations are based on the two-fluid formulation for CR transport
(\cite{dru81}). 
This method is much simpler than the more complete diffusion-convection
formalism of CR transport, and it is still the only really practical
method for studying multidimensional, time-dependent CR modified flows.
Although the two-fluid method has limits (\eg \cite{jonel91}), recent
work has demonstrated that the two-fluid formalism 
gives consistent solutions with diffusion-convection formalism
in time dependent shocks when the necessary closure parameters are
carefully defined (\eg \cite{kjr92}; \cite{fran95}; \cite{kanj95};
\cite{kanj96}).  

The outline of this paper is as follows. The detailed
description of the density spike formation is given in \S 2.  In \S 3,
we study the formation and evolution of the
density spike in a various shock structures by means of
one-dimensional numerical simulations.  The
modified criterion of Rayleigh-Taylor instability in the presence of
CRs is discussed by solving the linear perturbation equations of the
two-fluid system in \S 4.  \S 5 presents our two-dimensional numerical
results of the unstable density spike simulation, and we summarize in
\S 6.

\section{Formation of the Density Spike}

  Formation of the sharp density enhancement in CR modified shocks has been
explained previously (\cite{dru87,jon90}).  We
describe the process of density enhancement within CR-modified
shocks more fully here in order to understand better what conditions may
be relevant to the new instability we report.  The sharp density
enhancement or  ``spike'' 
originates from development of or increases in the CR precursor.
Diffusive acceleration at shocks depends on the ability of a fraction of
high energy
particles  isotropized by scattering in the postshock flow to return 
upstream of the shock. Most of those particles
are then advected back through the shock in a process that repeats. On
balance these two effects confine 
the upstream CRs within a region whose scale is the diffusion length,
$x_d \approx \kappa/v_0$,  
where $\kappa$ is a suitably defined mean spatial diffusion
coefficient (see \cite{dru81}), 
and $v_0$ is the upstream flow speed of the gas relative to the shock.
If we define $x_d$ in terms of the cosmic ray pressure, $P_c$, and
the CR pressure  increase before the shock is $\delta P_c$, then there
is a pressure gradient 
in front of the shock, $\sim \delta P_c/x_d$. That, in turn, decelerates the
gas flowing into the shock, which produces an adiabatic compression and heating
of the gas, adding to the total pressure gradient. 

For strong,
high Mach number and
steady shocks this ``precompression'' and ``preheating'' can be
sufficient to eliminate 
the need for a dissipative, gas subshock in order to satisfy the
steady jump conditions for the shock (\eg \cite{dru81}; \cite{malv96}). Then,
according to this theory, the steady
transition
between the upstream and downstream states will be continuous (a C-shock
is produced). On the other hand such a shock will usually form from one
that initially is not dominated by CR pressure. In other circumstances
the Mach number of the shock may increase, for example, as the shock
encounters a denser, colder environment or as the shock accelerates down
a steep density gradient. 
That shock strengthening is followed by an analogous increase in the
CR pressure and 
readjustment of the shock structure.
In either of these two situations
even an eventual C-shock will pass through an intermediate condition with a
precursor {\it and} a gas subshock.
In this transitional stage, the fluid is compressed twice; once by the
precursor and 
once by the gas subshock. 
The precompression of the gas within the precursor automatically reduces
the Mach number of the subshock.
However, if the subshock is initially strong, then while the CR
pressure at the shock is small compared to $\rho_0v_0^2$, 
the second density compression, through the subshock, is hardly
reduced from its initial value.
Consequently, the double transition of the flow can
result in a density compression of the flow much more than the
usual $\frac{\gamma_g+1}{\gamma_g-1}$ times the
density far upstream, where $\gamma_g$ is the adiabatic index of the gas.

In order to determine approximately 
the maximum possible compression ratio without solving the full set
of equations we can just use the above double compression concept.
Let us define four distinct states in the flow for convenience. 
State ``0'' refers to flow conditions far upstream, ahead of the
precursor. State ``1'' is that just upstream of the gas subshock, if
it exists, while state ``2'' is that just downstream of the subshock. Then,
finally ``3'' refers to the state that existed downstream of the
subshock before the shock began adjusting to new upstream conditions. 
That region will propagate downstream, away from the subshock, and it is
in the space between states ``2'' and ``3'' that the density spike
feature develops.
We also define the compression ratio through the precursor as
$D\equiv {\rho_1 \over \rho_0}$.
Recall that in this smooth transition the gas will be
compressed adiabatically, so that $P_g \propto \rho^{\gamma_g}$.
Then, one can use the gas adiabat
and, assuming that the compression through the precursor is slowly
changing,
mass flux conservation to obtain the
relation between the sonic Mach numbers (${\cal M}$) in states ``0'' and ``1''.
\begin{equation}
{\cal M}_1 = {v_1 \over c_1} = {v_0 \over c_0}{1 \over D^{{\gamma_g +1}
\over 2}} = {{\cal M}_0 \over D^{{\gamma_g + 1} \over 2}}
\end{equation}
where $c$ is the gas sound speed.  Therefore, we can see that the sonic
Mach number of the flow decreases as the flow
experiences the adiabatic compression.
If $P_{c3}$ is less than the equilibrium value, then $P_{c1}=P_{c0}+\delta P_c$ will
increase, causing  $D$ to rise as well.
The value of $D$ at equilibrium in a given shock depends on various details
such as the Mach number of the shock, but, from conservation laws
cannot exceed $D = D_{max} = 4$ if the net adiabatic index of the combined
thermal/CR plasma is nonrelativistic or $D = D_{max}= 7$ if that plasma is relativistic.
These limits would correspond to strong shocks that
have been smoothed entirely into C-shocks by CR pressure (\cite{dru81}).

As $D$ increases for a given ${\cal M}_0$ the value of ${\cal M}_1$
will decrease, down to a limit ${\cal M}_1 = 1$. If the entire
steady-state jump conditions can be satisfied when ${\cal M}_2 \ge 1$
the shock should evolve into a C-shock in which the downstream flow
is supersonic (\cite{dru81}). If the steady jump conditions across
the full transition require a subsonic flow in region ``2'', then a
subshock is required even as the shock reaches equilibrium.
Applying the condition ${\cal M}_1 = {\cal M}_2 \ge 1$
for $D = D_{max}$ leads to a minimum constraint on the Mach number
${\cal M}_0$ for a CR shock to be able to evolve into a C-shock; namely

\begin{equation}
{\cal M}_0 > D^{{\gamma_g + 1 \over 2}}_{max}.
\end{equation}
The critical Mach number for a C-shock lies in the range 6.35 to 13.39 for
$D_{max} $ between 4 and 7; that is as the net adiabatic index changes from
$\gamma_c = 5/3$ to $\gamma_c = 4/3$.  It should be noted that this
critical sonic Mach number is approximate because $D$ in C-shocks is
in general slightly
smaller than $D_{max}$. Therefore, the critical sonic Mach number for
C-shock should be slightly smaller than our estimated value.
The critical sonic Mach number increases for perpendicular shocks
when the magnetic field is strong
(\cite{jcn94}).  In the perpendicular shock case, the condition for a C-shock
is $v_1^2 (= v_2^2) > c_1^2 (= c_2^2) + v_{A,1}^2 (= v_{A,2}^2)$ where
$v_A$ is the Alfv\'en speed.  This 
gives the critical sonic Mach number
\begin{equation}
{\cal M}_0 > {(D^{\gamma_g + 1}_{max} + {v_{A,0}^2 \over c_0^2}D^3_{max})}^{1/2}.
\end{equation}

As long as the gas subshock remains, we can simply relate the states
``1`` and ``2'', by the standard shock jump conditions. An assumption 
in diffusive shock theory is that there is no discontinuity in $P_c$, since
the CRs necessarily are able to cross the subshock freely.
If there is a subshock, we can write
\begin{equation}
{\rho_2 \over \rho_1} = {(\gamma_g +1){\cal M}_1^2 \over (\gamma_g
-1){\cal M}_1^2 + 2} = {(\gamma_g +1){\cal M}_0^2 \over ({(\gamma_g
-1){\cal M}_0^2 \over D^{\gamma_g + 1}} + 2)D^{\gamma_g +1}}.
\end{equation}
If the subshock has vanished $\rho_2 = \rho_1$.
Including a subshock the total density jump between states ``0'' and ``2'' is then
\begin{equation}
{\rho_2 \over \rho_0} = {(\gamma_g +1){\cal M}_0^2 \over ({(\gamma_g
-1){\cal M}_0^2 \over D^{\gamma_g + 1}} + 2) D^{- \gamma_g }}.
\end{equation}
If we assume that the Mach number ${\cal M}_0$ is large compared to the
critical Mach number in equation 2, but there is still a strong subshock equation
5 reduces to
\begin{equation}
{\rho_2 \over \rho_0} \approx {\gamma_g +1 \over \gamma_g -1}D_{max}.
\end{equation}
This last expression cannot apply to a steady shock, but can be approximately
true during the evolution of the shock towards its time asymptotic 
state. For a relatively brief time compared to the evolution time scale
of the shock it can even happen that the subshock is strong (so that
${\cal M}_1 \rightarrow \infty$ in equation 4 but that $D\approx D_{max}$). Then the total
compression can exceed that of either the precursor or the subshock
separately.
In fact it can lead to compression as high as 28 (16) for $\gamma_c =
4/3$  ($\gamma_c = 5/3$) just as the shock approaches an equilibrium
and the spike detaches (the result indicated by equation 6). 
The extra compression is also accompanied by a modestly enhanced total
velocity jump 
during this time (\eg \cite{jon92}). This, in turn,
can increase the rate of particle acceleration,
at least for high energy particles whose scattering lengths are large
enough to allow them to cross substantial parts of the precursor
each time they pass through the subshock (\cite{dru87}).
This influence can speed up the modification process itself
as seen in Fig.5 of \cite{jcn94}.   

The overcompression forms a spike, because $\rho_3$, if it represents
an equilibrium condition for the earlier shock, will satisfy the
constraint, $r_3 = \rho_3/\rho_0 \le D_{max}$. Hence, for the
reasons we just discussed, there is a time when
$\rho_2  = r_2 \rho_0 > \rho_3$.
We can estimate the width of the spike through a simple
consideration of that transition interval. If the time to reestablish equilibrium is
$t_{eq}$, then the width of the spike is $x_s = \bar v_2 t_{eq}$, where
$\bar v_2$ is the mean velocity relative to the shock inside the
density spike. 
We can assume that the density $\rho_2$ peaks in time just before the
shock reaches its steady structure and define the total compression at that 
time to be $r_{2p} = \rho_2(peak)/\rho_0$.
Simplifying the density structure of the spike to a linear profile 
we can estimate $\bar v_2 \approx \frac{2 v_0}{r_3 + r_{2p}}$.
Then we can use a simple analytic estimate for $t_{eq}$ given 
by Jones \& Kang (1990) in terms of
the upstream diffusion time, $t_d = x_d/u_0 = \bar \kappa/u_0^2$, the
ratio, $P_{c2}/P_{c0}$ of the equilibrium downstream and upstream
cosmic-ray pressures and $\gamma_c$; namely,
\begin{equation}
t_{eq} \approx \frac{5 t_d}{g} \ln{[\frac{g(P_{c2}/P_{c0}) + 1 }{g + 1}]}, 
\label{teq}
\end{equation}
where $g = \frac{3 }{4}\gamma_c~-~1$.
In the limit $g \rightarrow 0$, $t_{eq} \approx 5 t_d (P_{c2}/P_{c0} - 1)$.
This leads to a simple estimate
for the width of the density spike at ``separation''; namely,
\begin{equation}
x_s \approx 10 \frac{x_d}{g(r_3+r_{2p})}  \ln{[\frac{g (P_{c2}/P_{c0}) + 1}{g + 1}]},
\label{xspike}
\end{equation}
with an appropriate limit taken as $g \rightarrow 0$. 
For moderate to strong shocks this formula will generally predict $x_s/x_d \sim$ few $\times (10 - 100)$, with larger
values for larger $P_{c2}/P_{c0}$ and as $\gamma_c\rightarrow 4/3$.

\section{One-Dimensional Numerical Simulations}

In order to study the formation process and evolution of the density
spike in CR-modified shocks, we use the modified-ZEUS code which
solves two-fluid equations on an Eulerian mesh.  The algorithms of the
code are described in Jun, Clarke, and Norman (1994).

Figure 1 shows the evolution of CR-modified shocks and density spikes computed
for different total Mach numbers defined by $v_0 / \sqrt{
{\gamma_g P_g \over \rho} + {\gamma_c P_c \over \rho}}$.   
To simplify our discussion we have chosen a constant diffusion coefficient,
$\kappa = 1$ for the moment. The value of the diffusion coefficient enters primarily
through its role in determining the width of the density spike
(equation 8) and through its role in controlling the
rate at which the CR modifications to the shock take place.  
Figs. 1a, 1b, and 1c show the evolution of CR-modified shocks with Mach
number 10, 20, and 50 respectively.   
The adiabatic index for CRs is $\gamma_c = 5/3$, for these particular
tests. Other values will be included later.
A shock wave is initially generated
by supersonic fluid striking a reflecting boundary at $x = 0$ and
moves to the right. The dotted lines represent the shock structure at
the earliest stage while the later stages are represented  by the
dashed line and solid lines in a time sequence.
The density overshoot is clearly seen in early stages of evolution.  
This density overshoot results in a ``density spike'' as 
the shock evolves further and the gas subshock becomes weaker
due to the decreased gas pressure which is replaced by CR pressure.
The shock front is strongly
modified by the CRs and the gas subshock disappears at later stage of
the evolution so that the shock transition becomes continuous (C-shock).
The formation of the density spike occurs as the density overshoot lags
behind the shock.
This lagging starts after temporary high pressure
(overshoot) from the shock front 
generates a positive total pressure gradient and starts to push the
density spike downstream.
As the overcompressed density spike moves downstream, the shock
structure reaches a steady state that can be determined by
conservation equations (\cite{dru81,web83,jcn94}).
The density spike is a moving compression
region, and material flows through it. We point out that
this feature is {\it not} a contact discontinuity.
The total compression ratio of the density spike is lower than
16 (for $\gamma_c = 5/3$) when it is formed at the shock front.
However, after the density spike moves downstream, the
compression continues and increases the density further so that
it can become higher than 16 for $\gamma_g = \gamma_c = 5/3$ (see Fig.1c).
The flow compression eventually slows down due to
the finite pressure gradient and increasing density.
We find in the overcompressed region that the density becomes higher
in a stronger shock (higher Mach number).  
The development of the density spike is
not restricted to the piston driven shock problem.  Figs. 1d, 1e, and 1f show
the evolution of CR-modified shocks which were initialized as a
standing shock by assigning analytic values for upstream
and downstream states.   Total pressure in the downstream is divided
equally to gas pressure and CR pressure.
The left boundary condition is outflow and
the right boundary is inflow with an upstream condition. The results
show qualitatively the same density spike evolution as for the
CR-modified piston driven shocks.   Once detached, the density
spike appears to move leftward since the shock front is at rest.
The density spike becomes sharper as Mach number increases because the
diffusion length is inversely proportional to the shock speed (see
equation 8).

In Fig.2, we present the evolution of three different CR-modified
shocks.  Fig.2a shows the formation of the density spike in a Mach 20
standing shock with the CR adiabatic index $\gamma_c = 4/3$.  The shock
transition becomes continuous eventually due to the CR domination in
the downstream state of the shock.  This CR domination of the shock
also results in the steady-state density jump of the shock higher than 4.   The
prominent density spike is also found in the downstream of the shock
as expected.  Fig.2b shows the result of CR-modified shock with the
diffusion coefficient $ \kappa = 1/\rho$.  The rest of initial conditions are
identical to the case presented in Fig.1d.  As reported by Jun et
al. (1994), the steady state is obtained faster than in the case with
$\kappa =1$.  Another feature is the sharper structure of the shock
and density spike because of the smaller diffusion length in the shock
transition region and downstream state where the density is higher than 1.  
Fig.2c shows the evolution of CR-modified shock which contains a gas
subshock and a smooth precursor.  The density spike is also clearly present,
although it is not as prominent as in C-shock case.  This
example manifests that the density spike is not restricted to the
C-shock case.   This density
enhancement always follows the formation or enhancement of a  precursor.  Therefore, the
density augmentation should form whenever the CR acceleration is
efficient and the back-reaction of CR to the shock front is
dynamically important. 

The density spike also forms in simulations involving shock propagation
into nonuniform media. It has been seen in cases where the shock
moves down a steep density gradient, such as a SNR blast into a
preexisting wind cavity, or
where a CR-modified shock encounters a steep increase in upstream
density, such as a cloud.  Fig.3 shows the evolution of a Mach
10 shock propagating into a higher density background.  The shock is
initialized as a standing shock at x=3.  The shock is already modified
by generating a precursor and density overshoot at t=0.1.   By t=0.2,
a density spike has formed and is moving to the left relative to the
shock front.   The shock transition has
reached a steady state and the gas subshock disappeared
(leaving a C-shock).  At t=0.3, the C-shock is approaching a higher
density medium ($\rho = 3$), while at t=0.4, the C-shock is interacting with the
higher density region and another density overshoot is produced. 
Finally, at t=0.5, the shock reached to
the new steady state and the second density spike is formed.  The
C-shock structure is recovered (Figs. 3e and 3f).

As the CR-modified shock encounters the denser external medium, 
its precursor is compressed 
and the shock structure temporarily reforms a viscous, gas subshock.
This process can occur if 
the rate of increase of the incoming gas density exceeds the
rate at which the shock is able to adjust to an equilibrium in terms of its
CR pressure, as given by equation 7.
As the newly formed
discontinuous shock propagates into the denser background, the
shock is modified anew and the precursor is enhanced.  The important effect
during the shock passage into the denser region is a higher
acceleration efficiency due to increased compression. 
If the inhanced upstream density is in pressure equilibrium with
its surroundings, it will be colder. Then it is easy to show that the
Mach number of the shock increases as it penetrates the denser medium.
The importance of this will
be more significant in an already CR-dominated C-shock, because the
full shock
compression can become much higher in responce to the density
enhancement encounter than through the original incoming C-shock.
Jones and Kang (1992) found in their shock-cloud two-fluid simulations 
that the acceleration efficiency is increased more by the encounter
if the incident shock is previously dominated by CR pressure.

In the begining of the density spike formation, the total pressure
decreases upstream of the density spike in order to conserve total momentum
flux ($\rho v^2 + P_g + P_c$).  As a 
result, the total pressure in the left side of the density spike has a
negative gradient while the density gradient is positive.  As we will
show in the next section, these opposite gradients between the total
pressure and the density consitute an unstable condition.  
As the shock structure itself reaches a dynamical steady state (equilibrium), the
momentum flux upstream of the density spike becomes constant
and the steepening of the density spike slows down.
Eventually, the gradients of the total pressure and the density decreases, after the
density spike is detached from the shock front, mainly because of CR
diffusion.  Once this happens, the growth rate of the instability decreases
accordingly.   From this dynamical picture it is obvious
that the duration of the instability is comparable to the
shock equilibrium timescale (equation 7) each time the shock readjusts.
Since the opposite gradients of the total pressure and the density
exist only for a short period of time during the formation of
the density spike, the driving force of the instability appears to be
impulsive. 

\section{Modified Rayleigh-Taylor Instability}

 One can expect that cosmic-ray pressure should be added to the
total pressure in the criterion for the Rayleigh-Taylor instability
in the presence of cosmic-ray particles.   In this section, we derive
the modified criterion for the Rayleigh-Taylor instability in the
presence of CR pressure by considering the two-fluid model
(\cite{dru81}). The governing equations are:
\begin{equation}
{\partial \rho \over \partial t} + \nabla \cdot (\rho {\vec v}) = 0
\end{equation}
\begin{equation}
\rho {\partial {\vec v} \over \partial t} + \rho({\vec v} \cdot \nabla
){\vec v} +  \nabla (P_g  + P_c) + \rho {\vec g} =0
\end{equation}
\begin{equation}
{\partial P_g \over \partial t} + {\vec v} \cdot \nabla P_g + \gamma_g
P_g \nabla \cdot {\vec v} = 0
\end{equation}
\begin{equation}
{\partial P_c \over \partial t} + {\vec v} \cdot \nabla P_c + \gamma_c
P_c \nabla \cdot {\vec v} = \nabla \cdot (\kappa \nabla P_c),
\end{equation}
where $P_c$ is the CR pressure and $\kappa$ is the mean spatial CR diffusion
coefficient.
The gravity term is added to the momentum equation to satisfy the
equilibrium state in a stationary condition.
For this analytical analysis we assume $\kappa$ to be a constant, but that is
not necessary, either for this instability, or for validity of
the two-fluid model (see \cite{kjr92};~ \cite{kanj95};~\cite{kanj96} for specific
tests of time-dependent two-fluid simulations against more detailed
diffusion-convection equation or ``particle'' simulations).
We consider an exponential background,
\begin{equation}
\bar{v_x} = v_0 exp({x-x_0 \over L_1}), \quad
\bar{\rho} = \rho_0 exp({x-x_0 \over L_2}), \quad
\bar{P_g} = P_{g0} exp({x-x_0 \over L_3}), \quad
\bar{P_c} = P_{c0} exp({x-x_0 \over L_4}).
\end{equation}
In general, the normal mode for a nonuniform background contains a  
dependence on the x-direction, and one needs to solve the characteristic
value problem with proper boundary conditions.
However, since we are seeking the instability criterion only, 
we can  restrict our analysis to the local region, $x-x_0 \ll L_1$.
We consider plane wave perturbations, $\sim \psi_1 exp(nt
+ ik_yy)$ where $\psi_1$ is a perturbed quantity ($v_{x1}$, $v_{y1}$,
$\rho_1$, $P_{g1}$, or $P_{c1}$). 
Then, one can obtain the linearized perturbation equations using equations 9 - 13,
\begin{equation}
\rho_1(n + {v_0 \over L_1}) + ik_y \rho_0 v_{y1} + {\rho_0 \over L_2}
v_{x1} = 0,
\end{equation}
\begin{equation}
v_{x1} (n+ {v_0 \over L_1}) - {\rho_1 \over \rho_0^2}({P_{g0} \over
L_3} + {P_{c0} \over L_4}) = 0,
\end{equation}
\begin{equation}
nv_{y1} + {ik_y \over \rho_0} (P_{g1} + P_{c1}) = 0,
\end{equation}
\begin{equation}
P_{g1}(n + \gamma_g {v_0 \over L_1}) + v_{x1}{P_{g0} \over L_3} + ik_y
\gamma_g P_{g0} v_{y1} = 0,
\end{equation}
\begin{equation}
P_{c1}(n + \gamma_c {v_0 \over L_1} + \kappa k_y^2) + v_{x1}{P_{c0}
\over L_4} + ik_y \gamma_c P_{c0} v_{y1} = 0.
\end{equation}
By combining the above equations, we get the dispersion relation,
\begin{eqnarray}
&& \left( \rho_0(n + {v_0 \over L_1})^2 + {1 \over L_2}( {P_{g0} \over
L_3} + {P_{c0} \over L_4}) \right) \nonumber
\\
& \times & \left(n(n + {\gamma_g v_0 
\over L_1})(n + {\gamma_c v_0 \over L_1} + \kappa k_y^2) + k_y^2
a_g^2(n + {\gamma_c v_0 \over L_1} + \kappa k_y^2)
+ k_y^2a_c^2(n + {\gamma_gv_0 \over L_1})\right) \nonumber \\
& -&  {k_y^2 P_{g0} \over \rho_0L_3}(n + {\gamma_c
v_0 \over L_1} + \kappa k_y^2)( {P_{g0} \over L_3} + {P_{c0} \over
L_4})  - {k_y^2P_{c0} \over \rho_0 L_4}(n
+ {\gamma_g v_0 \over L_1})( {P_{g0} \over L_3} + {P_{c0} \over L_4}) = 0
\end{eqnarray}
where $a_g = ({\gamma_g P_{g0} \over 
\rho_0})^{1/2}$, and $a_c = ({\gamma_c P_{c0} \over \rho_0})^{1/2}$.

   Let  us consider two different limits of this dispersion
relation in a stationary background ($v_0 \rightarrow 0$).

\subsection{Short-Wavelength Limit ($k_y \rightarrow \infty$)}

The familiar and related Rayleigh-Taylor and convective instabilities in pure gas
dynamics are actually somewhat different in terms of criteria, because
the convective instability depends on fluid compressibility, whereas
the classical Rayleigh-Taylor instability does not.  
A flow is convectively unstable if $ {\partial
ln \bar{\rho} \over \partial x}/{\partial ln \bar{P_g} \over \partial
x} < 1/{\gamma_g}$, while the flow is Rayleigh-Taylor unstable if  $ {\partial
ln \bar{\rho} \over \partial x}/{\partial ln \bar{P_g} \over \partial
x} < 0 $ (\cite{ban84}). This shows that the convective
instability is present in a wider range of conditions than the
Rayleigh-Taylor instability. 
When the CR diffusion term is ignored, our dispersion relation reduces
to the growth rate for the general convective instability in the
short-wavelength limit (for the frequency range, $n \gg \kappa k_y^2$ but
$n^2 \ll k_y^2(a_g^2 + a_c^2)$). 
\begin{equation}
n^2 = {1 \over \rho_0^2(a_g^2 + a_c^2)}({P_{g0} \over L_3} + {P_{c0}
\over L_4})^2 - {1 \over \rho_0 L_2} ({P_{g0} \over L_3} + {P_{c0}
\over L_4}) \simeq {1 \over \gamma_t \rho_0}({\partial \bar{P_t} \over
\partial x})({\partial ln(\bar{P_t}/\bar{\rho}^{\gamma_t}) \over \partial x}).
\end{equation}
where $\gamma_t = {\gamma_gP_{g0} + \gamma_cP_{c0} \over P_{g0} +
P_{c0}}$ and $\bar{P_t} = \bar{P_g} + \bar{P_c}$.

In the presence of CR diffusion and short wavelength limit (for the
frequency range, $n \ll \kappa k_y^2$ and $n^2 \ll k_y^2(a_g^2 +
a_c^2$), 
the dispersion relation reduces to 
\begin{equation}
n^2 = {1 \over \rho_0}({P_{g0} \over L_3} + {P_{c0} \over L_4})({1
\over \gamma_g L_3} - {1 \over L_2}) \simeq {1 \over \rho_0}{\partial
\bar{P_t} \over \partial x}({1 \over \gamma_g }{\partial
ln \bar{P_g} \over \partial x} - {\partial ln \bar{\rho} \over
\partial x}).
\end{equation}
This relation includes criteria for both gas convection and
Rayleigh-Taylor instabilties.
Now, let us  consider our dispersion relation for several different cases.

 i) First, when all three gradients ($\partial \bar{P_t}/ \partial x, \partial
\bar{P_g} / \partial x$, and $\partial \bar{\rho} /\partial x$) have same
sign, then the flow can be convectively unstable if
$ \vert{1 \over \gamma_g}{\partial ln \bar{P_g} \over
\partial x}\vert > \vert{\partial ln \bar{\rho} \over \partial x}\vert$.

  ii) Second, the flow is always Rayleigh-Taylor unstable if 
\begin{equation}
{\partial \bar{P_t} \over \partial x} > 0,  {\partial \bar{P_g} \over
\partial x} >0,  {\partial \bar{\rho} \over \partial x} <0
, \quad or \quad
{\partial \bar{P_t} \over \partial x} < 0,  {\partial \bar{P_g} \over
\partial x} <0,  {\partial \bar{\rho} \over \partial x} >0.
\end{equation}

  iii) Third, the flow is Rayleigh-Taylor unstable if 
\begin{equation}
{\partial \bar{P_t} \over \partial x} > 0,  {\partial \bar{P_g} \over
\partial x} <0,  {\partial \bar{\rho} \over \partial x} <0
, \quad or \quad
{\partial \bar{P_t} \over \partial x} < 0,  {\partial \bar{P_g} \over
\partial x} >0,  {\partial \bar{\rho} \over \partial x} >0,
\end{equation}
when the gas is convectively stable ($ \vert{1 \over
\gamma_g}{\partial ln \bar{P_g} \over \partial x}\vert <
\vert{\partial ln \bar{\rho} \over \partial x}\vert $).  It is
interesting to see that the convective instability can be suppressed
by the opposite CR pressure gradient.

\subsection{Large-Wavelength Limit ($k_y \rightarrow 0$) }

In the large-wavelength limit (for the range, $n \gg \kappa k_y^2$ and
$n^2 \gg k_y^2(a_g^2 + a_c^2)$), the dispersion relation reduces to
\begin{equation}
n^2 = -{1 \over L_2 \rho_0}({P_{g0} \over L_3} + {P_{c0} \over L_4})
\simeq -{1\over \rho_0^2} {\partial \bar{\rho} \over \partial
x}{\partial (\bar{P_g} + \bar{P_c}) \over \partial x}.
\end{equation}
Therefore, the flow is unstable if
\begin{equation}
{\partial \bar{\rho} \over \partial x}{\partial \bar{P_t} \over
\partial x } < 0.
\end{equation}
This is the generalized crossed-gradient condition for Rayleigh-Taylor
instability when the total pressure is considered instead of the gas
pressure (e.g. see \cite{jss81}).    One can see that the convective
instability disappears in the large-wavelength limit.

  From the above analysis, one can see that the flow is Rayleigh-Taylor
unstable throughout the entire wavelength range when the total pressure
and gas pressure gradients are in opposite sign to the density gradient,
which is the instability condition in our density spike case.

\section{Two-Dimensional Numerical Results and Discussion}

The nonlinear development of the unstable density spike is
best studied by means of multi-dimensional
numerical simulations.  We have simulated the instability of the
density spike in two dimensions by setting up several different 
CR-modified shocks, such
as piston-driven shocks and standing shocks.  The density spike is found
to be unstable in every case. To reduce the potential that we are
observing only a numerical effect, we carried out analogous
simulations with two very different codes; namely a
version of ZEUS, modified to include two-fluid CR physics (\cite{jcn94})
and a version of PPM, also modified and used extensively in the past for
a number of CR applications (e.g., \cite{jon90};\cite{jon93};\cite{rkj93}). The
gasdynamical methods 
are quite distinct in the two codes, and the CR transport methods are
also different (one being explicit and one being implicit, for
example). 
The results of the two codes were qualitatively the same, and most
importantly, both demonstrated the instability.
Fig. 4 shows the two-dimensional results of a Mach
10 CR-modified shock structures computed by the modified ZEUS code. 
The one-dimensional time-dependent
evolution of the same shock is shown in Fig.1d.
The computational space is resolved by 600x200 uniform zones.  A random
density perturbation (10\%) is carried in by the flow from the right boundary.  The
density perturbation (pressure-free perturbation) is chosen because we
do not intend to generate sound waves that are unstable in the
CR-dominated shock (acoustic instability, see,
e.g., \cite{dru86};~\cite{kjr92}).  In this way, we should be free of
any influence from the propagating acoustic instability, so 
that the instability we observe in the density spike
results only from the Rayleigh-Taylor instability.

Images in Fig. 4 show the density at t=0.1, 0.2, and 0.3 (top to bottom).
A number of thin fingers are produced in the left side of the density
spike as a result of the instability.
In addition, these fingers show unstable curly structures that
are likely to be the result of the secondary Kelvin-Helmholtz
instability which was triggered by the shear between the finger and
the background flow downstream of the Rayleigh-Taylor unstable region.  
The evolution of the unstable flow
is found to be different from the classical Rayleigh-Taylor instability
in the following way.
In the classical Rayleigh-Taylor instability, 
short wavelength modes appear first in the linear regime, because their
growth rates are higher.
However, the nonlinear Rayleigh-Taylor instability is generally
dominated by larger wavelength features, because they reach higher
terminal velocity and overtake smaller structures (\cite{jns95}).
However, short wavelengths are still dominant in the nonlinear stage
of our CR-modified shock case.    This is because the
unstable condition exists only temporarily and does not provide
sufficient time for growth of perturbations on the scale of the thickness
of the density spike itself. (Recall that the perturbation wavevectors
are transverse to the flow direction.)
Short wavelength features are just stretched along the flow direction
by the initial acceleration to form long thin fingers
without much interaction with one another.

To provide a sense of the time evolution of the instability,
Fig.5 shows the amplitude of the density fluctuations as a function of
time.  The amplitude, A(x) is defined as ${1 \over H} \int_O^H \delta
\rho (x,y)^2 dy$ where $H$ is the width of the computational plane in
the Y-direction and $\delta \rho(x,y) = {\rho(x,y) \over
\bar{\rho(x)}} -1$.  This 
amplitude is averaged over the region $x_2 - x_1$ in the unstable
flow; that is, $A = {1 \over x_2 - x_1} \int_{x_1}^{x_2} A(x) dx$.  The
domain for the average is chosen to be 10 zones leftward from the peak
density in the density spike.  The amplitude, A provides us some
information about the overall density fluctuation history.  We find that
Fourier amplitude of the density perturbation is not
a good way to measure the growth of the instability because the flow
is moving through the unstable region. 

The apparent behavior of A(x) allows us to divide the growth history
of the instability into 4 different stages as indicated in Fig.5.  In
stage 1, the flow carrying density perturbation has not reached the density
spike.  The amplitude of the density perturbation increases suddenly
in stage 2 due to the incoming density perturbation. The density
perturbation grows while the flow with the density perturbation passes
through the unstable region (stage 3). Recall that the growth of the
instability exists 
only temporarily while the flow in the left (downstream) side of the
density spike is unstable.  The, final, relaxation phase of the instability is
shown in stage 4.  
The sudden decay of the instability growth is
likely due to the combined effects of relaxation of the unstable
region and the nonlinearity of the instability.
The instability is found
to result in a slight reduction in density in the spike compared
to the one-dimensional behavior at this time.  However, this reduction
is small because of the short duration of unstable flow at the early stage.

Although our perturbations were designed to avoid generation of
upstream sound waves that might contribute to this instability, we
recognized that passage of the density perturbations through the
shock might also produce sonic perturbations. To avoid any possibility
that this might modify the instability of interest,
we have done a simulation with the diffusion coefficient,
$\kappa = 1 / \rho$, because it is known that CR-modified
shocks are stable against the accoustic instability when $\kappa$
takes this form (\cite{dru86}).  Fig.6 shows a greyscale
image of a two-dimensional numerical result for the case with $\kappa = 1/\rho$
at t=0.2 (one-dimensional result for this case is presented in Fig.2b).
The density
spike is still unstable, thus, confirming the existence of
the physical Rayleigh-Taylor instability in the density spike.
The shorter density fingers compared to Fig.4b are due
to the narrower density spike caused by the reduction in
diffusion length and equilibrium time when $\kappa = 1/ \rho$
(compare Fig.1d and Fig.2b). In the presence of sonic perturbations
, \cite{rkj93} discovered in two-dimensions there
is a secondary, Rayleigh-Taylor class instability associated with
the nonlinear accoustic instability, which generates accoustic
turbulence within the shock precursor. Since that instability will
also act as strong CR-modified shocks come to equilibrium, it may be
natural to expect that the density spike will often be subject to strong
perturbations.

An obvious place where one can expect the unstable density spike to 
be formed may be in
the supernova remnant blast wave structure.   Numerical studies of
CR-modified supernova remnant shocks reveal the appearance of the
density spike in the evolution (\cite{jon92}).  Stronger modification
of the shock structure by CR pressure was found in the simulations
when the supernova remnant shock
propagates into an inverse square density medium such as a wind-blown
background.   Therefore, this density spike could be found in the
early stage of supernova remnant evolution while the supernova shock
is still inside the presupernova wind.  Discovery of this unstable
density spike and consequent turbulent flow would
provide strong, {\it in situ} evidence within supernova remnants for the
action of diffusive shock acceleration and, especially for strong shock
modification by energetic particles. We note finally that, although these
effects are strongest for C-shocks, the development of C-shocks is {\it not}
necessary for what we have discussed in this paper. It is necessary, however,
that CR pressure become the dominant means of decelerating the flow
through the shock.

\section{Summary}

  We have given a detailed description and numerical study
of the density spike formation
downstream of a CR-modified shock structure.  The density
spike originates from a temporary overcompression while the gas shock
is modified by the CR pressure.  The resulting compression ratio is
found to increase with the flow Mach number.    We emphasize that the
density spike is a common relic of the shock modification by the CR
pressure.  The density spike is
found to be Rayleigh-Taylor unstable at the early stages of its evolution.  
Our linear analysis shows that the flow is Rayleigh-Taylor unstable if the
total pressure gradient and gas density gradient are of opposite
signs.   We have simulated this unstable density spike by solving the
time dependent two-fluid equations numerically on a two-dimensional grid.
It is found that the instability grows impulsively in the early stages,
but slows down as the density spike begins to relax its structrure.  
The numerical results show a number of thin fingers
growing downstream from the density spike.  
However, the mass density in the
spike is decreased only slightly as a result of the weak instability.

\acknowledgements

The simulations were performed on a Cray C90 at the Minnesota Supercomputer
Center.   The work reported here is supported in part by NSF grant
AST-9318959 by NASA grants NAGW-2548 and NAG5-5055 and by the Minnesota Supercomputer Institute.

\clearpage

\figcaption[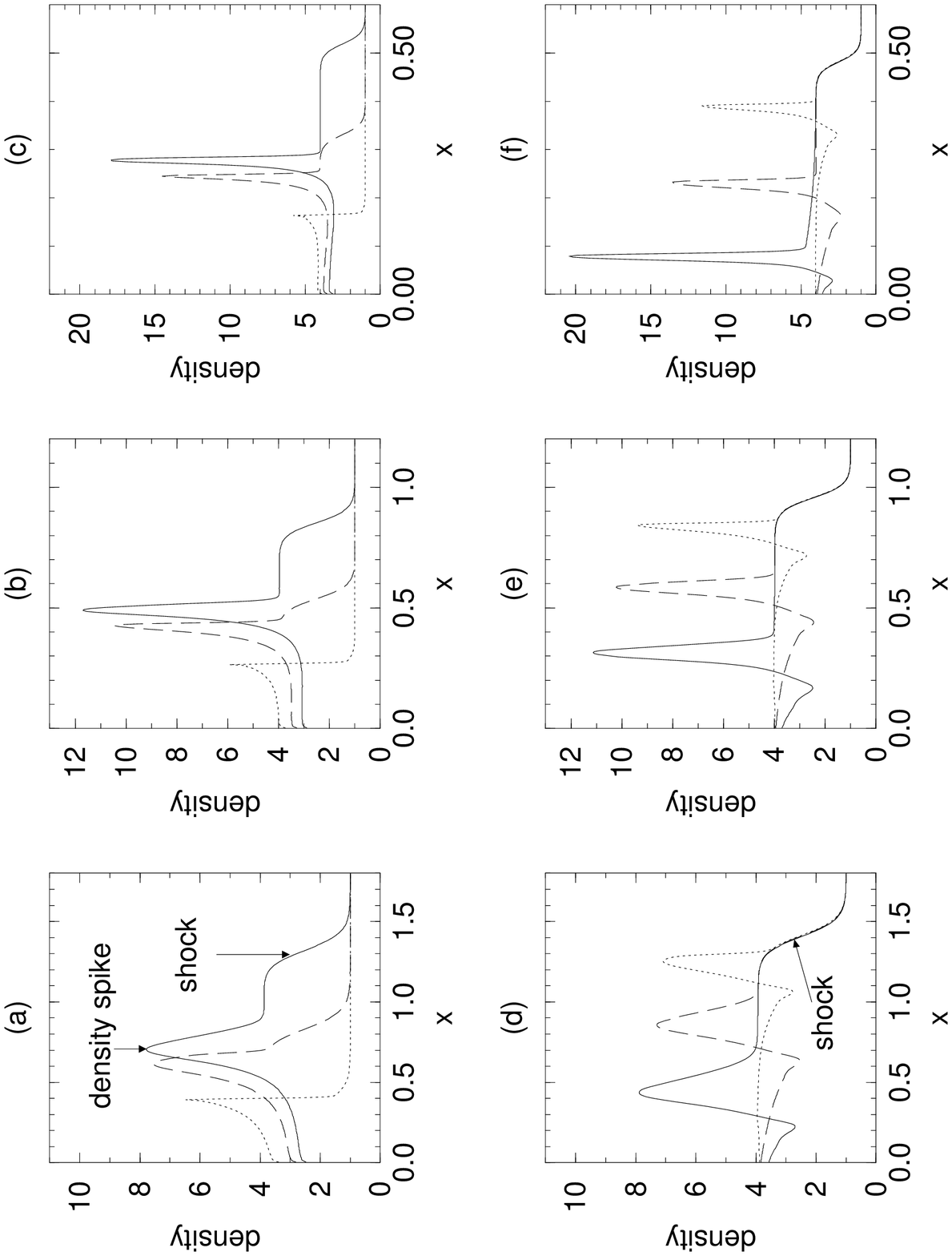]{ Density evolution of CR-modified shock structures for
various Mach numbers. The shock is propagating into a uniform background
of $\rho=1, P_g=1, P_c=1$.  Adiabatic index, $\gamma$, is assumed to be 5/3 for
both gas and CR and the diffusion coefficient, $\kappa = 1$.   Each plots (dotted
line, dashed line, and solid line) are represented in a time
sequence. The computational domain is resolved by 300 uniform grid
zones.  Readers are referred to Dorfi(1984,1985), Jones \& Kang(1990),
and Jun et al. (1994) for the evolution of other dynamical variables.
(a) Piston driven
shock with Mach 10. The shock is propagating with the velocity $v_s = 18.257$.
Three density plots are taken at t= 0.1,0.2, and
0.3. (b) Piston driven shock with Mach 20 ($v_s = 36.515$).  Three density
plots are taken at t= 0.033, 0.066, and 0.099. (c) Piston driven shock
with Mach 50 ($v_s = 91.287$). Three density plots are taken at
t=0.008, 0.016, and 0.024. (d) Standing shock with Mach 10 ($v_s =
18.257$).  The upstream ($\rho=1.0, v=-18.257, P_g=1.0, P_c=1.0$) and
downstream ($\rho=3.884, v=-4.7, P_g=124.75, P_c=124.75$) states of
the flow are initialized with  
analytic jump condition of CR-modified shock.  Total pressure in the
downstream is divided equally to gas pressure and CR pressure.
Three density plots are taken
at t=0.1,0.2, and 0.3. (e) Standing shock with Mach 20 (upstream
state: $\rho=1.0, v=-36.515, P_g=1.0, P_c=1.0$. downstream state:
$\rho=3.97, v=-9.197, P_g=499.75, P_c=499.75$).  Three density
plots are taken at t=0.033, 0.066, and 0.099. (f) Standing shock with
Mach 50 (upstream state: $\rho= 1.0, v_s = -91.287, P_g=1.0,
P_c=1.0$. downstream state: $\rho=3.995, v_s=-22.847, P_g=3124.7,
P_c=3124.7$ ). Three density plots are taken at t=0.008,
0.016, and 0.024. \label{fig1}}  

\figcaption[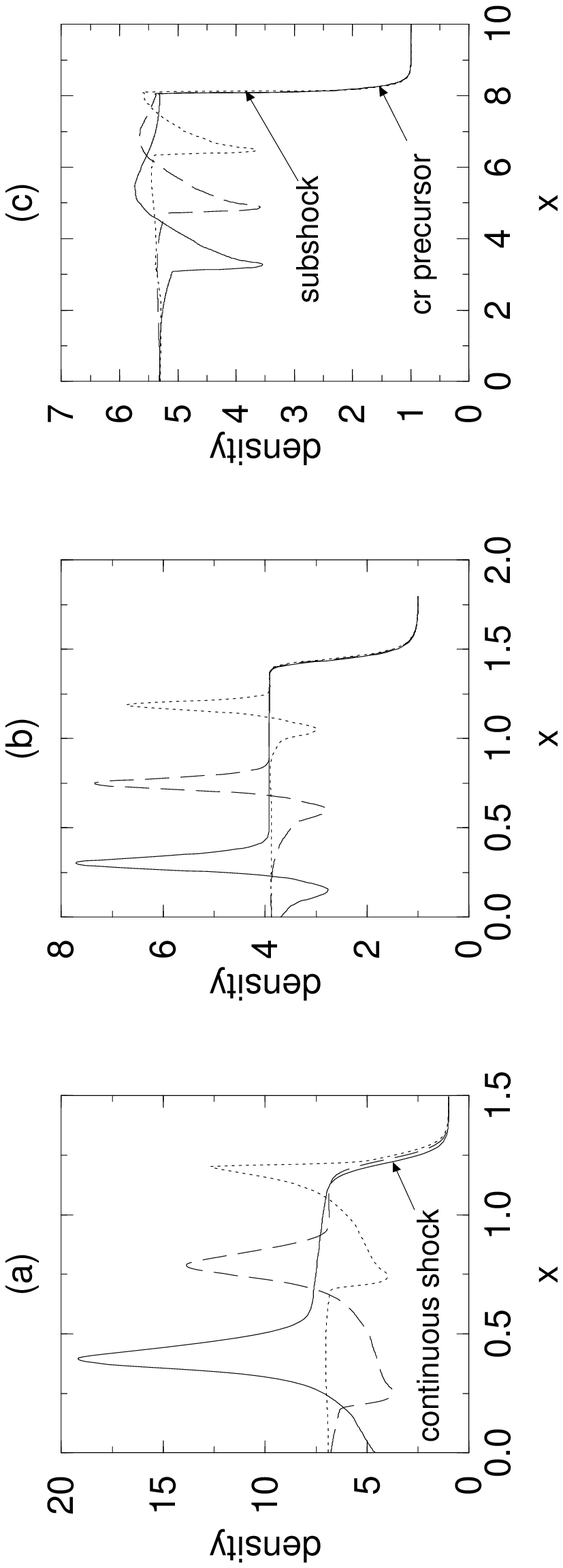]{ (a) Standing CR-modified shock (Mach 20, $v_s =
34.641$).  The adiabatic index for CR $\gamma_c = 4/3$ and $\kappa = 1$.
Plots are taken
at t=0.1 (dotted line), t=0.2(dashed line), and t=0.3(solid line).
The numerical resolution of the entire computational space is 300 zones.
(b) Standing
CR-modified shock (Mach 10) with the diffusion coefficient $\kappa =
1/\rho$.  Other physical conditions are identical to the case in Fig.1d.
(c) Standing CR-modified shock (Mach 5, $v_s = 8.66, \gamma_c = 4/3$).
The computational space is resolved by 800 grid zones. \label{fig2}}

\figcaption[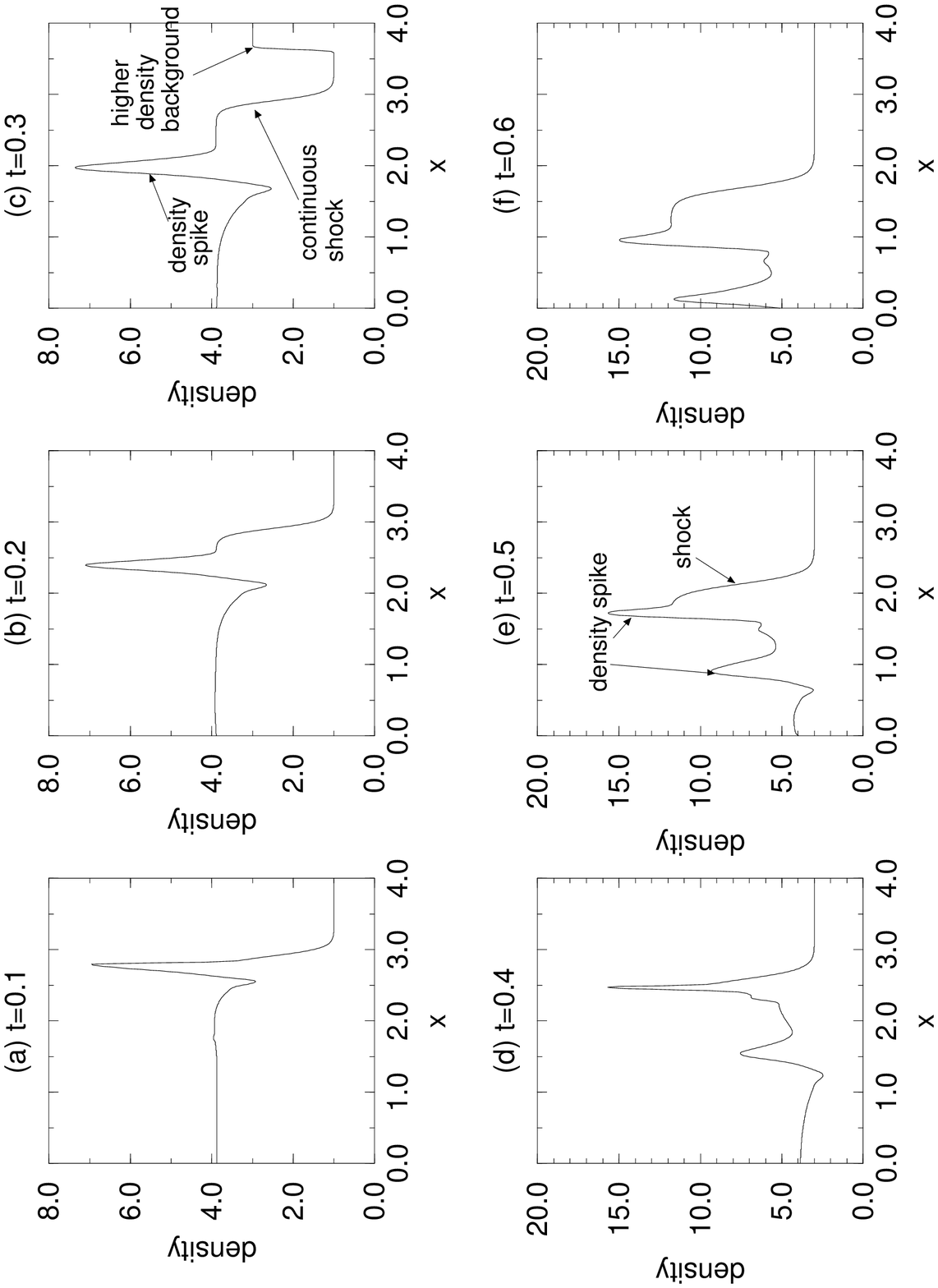]{Mach 10 shock ($v_s = 18.257$) is propagating
into the higher density region ($\rho = 3$). The adiabatic index for
CR is $\gamma_c = 5/3$, and $\kappa = 1$. The computational space is
resolved by 400 grid zones. \label{fig3}} 

\figcaption[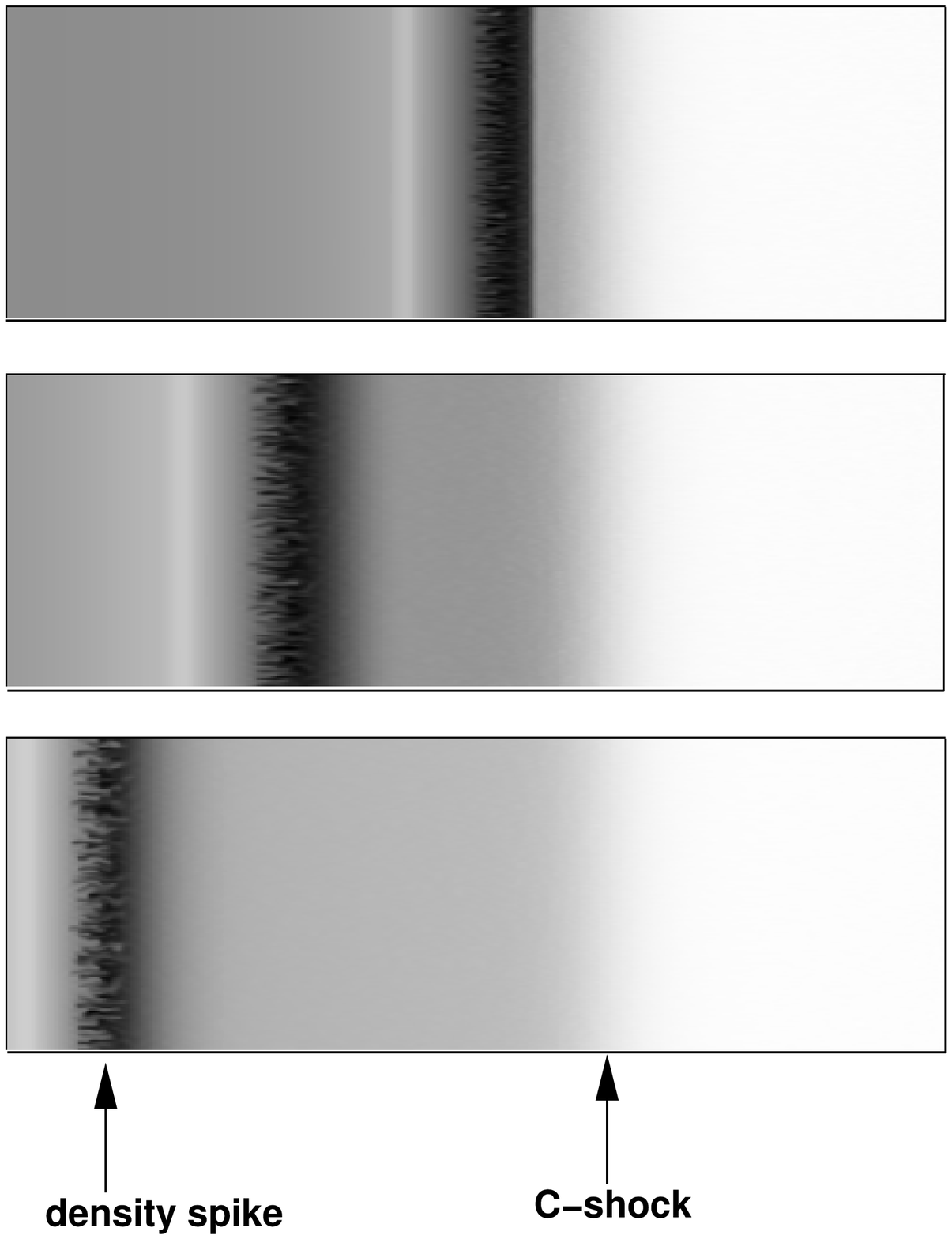]{ Grey-scale images show the density of two-dimensional
numerical result of CR-modified standing shock with Mach 10 at t=0.1, 0.2, and
0.3 (from top to bottom).  The computational space is resolved by
600x200 uniform grid zones.  One-dimensional result is shown in
Fig.1d. \label{fig4}} 

\figcaption[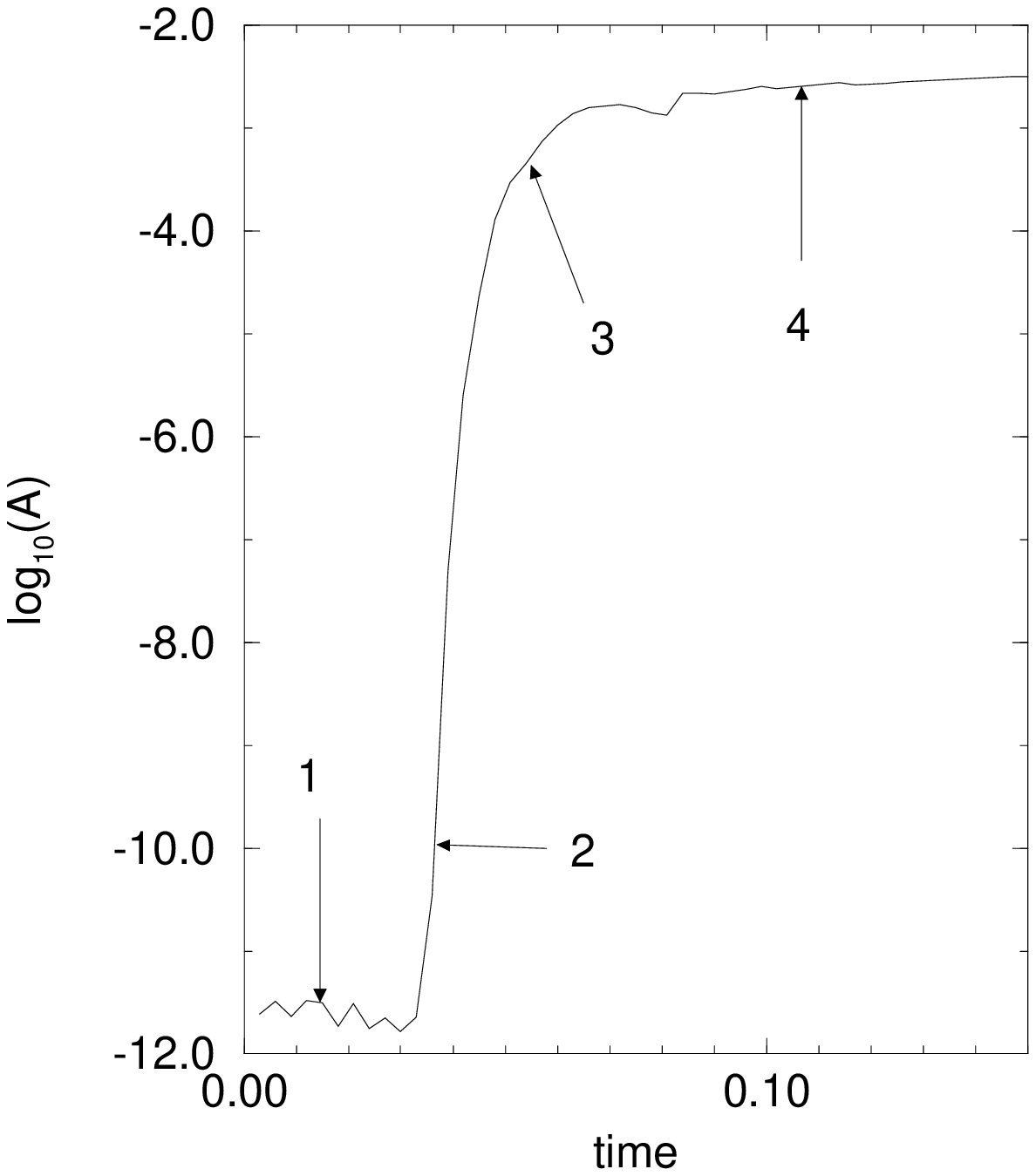]{ The growth of the density perturbation in the density
spike in Mach 10 standing shock (fig.4).  $$A(x) = {1 \over H}
\int_0^H \delta \rho (x,y)^2 dy$$ where 
$H$ is the width of the computational plane in Y-direction, $\delta
\rho (x,y) = {\rho(x,y) \over \bar{\rho(x)}} -1$, and $\bar{\rho(x)}$ is
the averaged 
density over Y-direction. $$A = {1 \over x_2 - x_1}
\int_{x_1}^{x_2} A(x) dx$$ where the average is taken over the region
$x_2 - x_1$ in the unstable flow.  \label{fig5}}

\figcaption[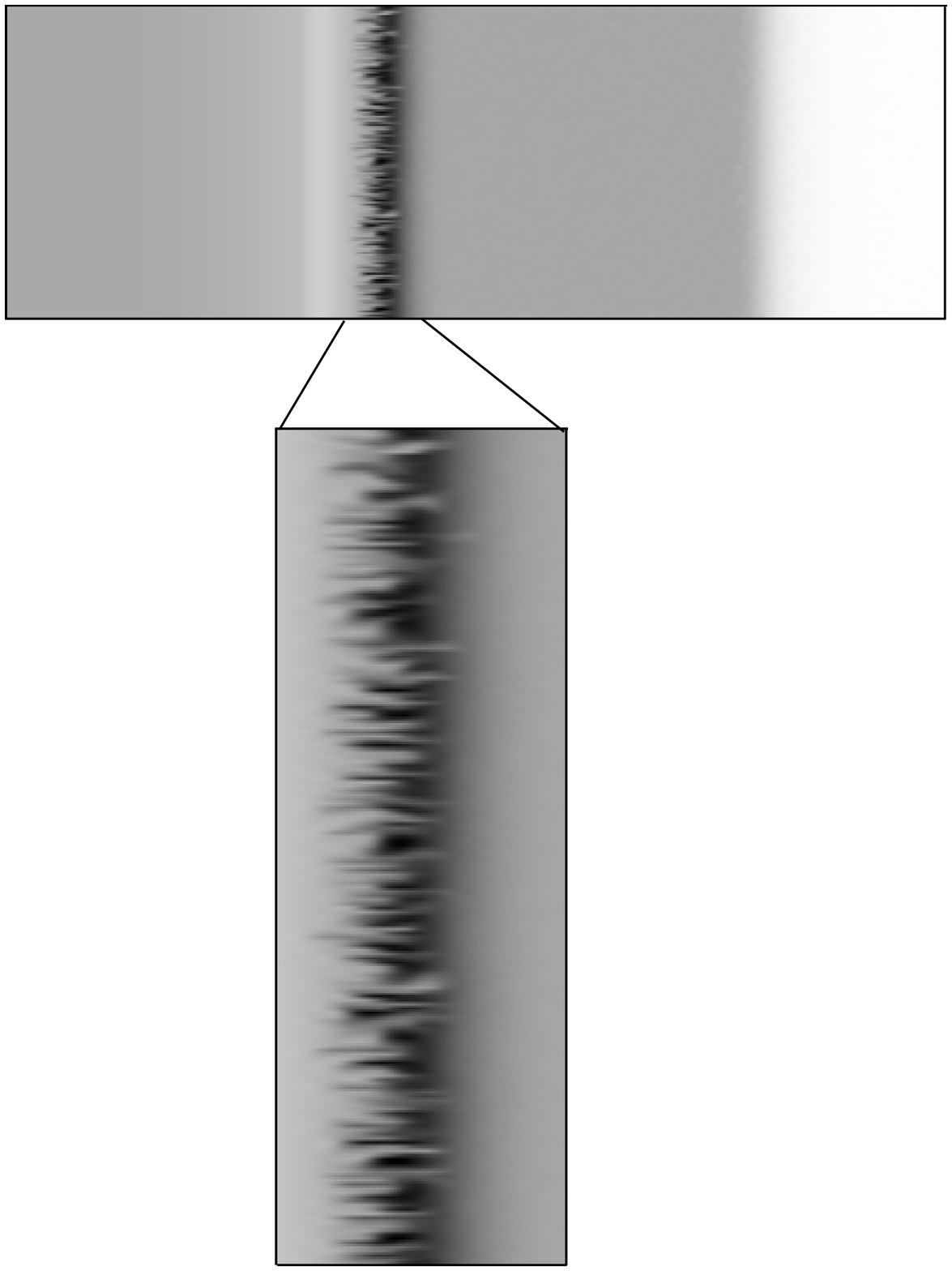]{ Grey scale image of CR-modified shock
generated by Mach 10 shock at t=0.2.    The diffusion coefficient used
is $1/\rho$. Other physical parameters are identical to the ones used
Fig.2b case.  The computational space is
resolved by 600x200 uniform grid zones. \label{fig6}}


\begin{thebibliography}{}

\bibitem[Axford et al., 1977]{als77} Axford, W.I., Leer, E., and
Skadron, G. 1977, Proc. 15th Internat. Cosmic Ray Conf(Plovdiv), 11, 132

\bibitem[Bandiera 1984]{ban84} Bandiera, R. 1984, \aap, 139, 368

\bibitem[Bell 1978]{bel78} Bell, A. R. 1978, \mnras, 182, 147

\bibitem[Blandford \& Ostriker 1978]{bla78} Blandford, R. D., \& 
Ostriker, J. P. 1978, \apj, 221, L29 

\bibitem[Blandford 1980]{bla80} Blandford, R. D. 1980, \apj, 238, 410

\bibitem[Dorfi 1984]{dor84} Dorfi, E. A. 1984, Adv. Space Res., 4, 205

\bibitem[Dorfi 1985]{dor85} Dorfi, E. A. 1985, in Cosmical Gas
Dynamics, ed. F. D. Kahn (Utrecht: VNU Science Press), p. 137

\bibitem[Drury 1987]{dru87} Drury, L.O'C. 1987, Proc. Sixth Int. Solar
Wind Conf., ed. U. J. Pizza, T. Holzer, and D. G. Sime
(NCAR/TN-360+proc), p.521 

\bibitem[Drury \& Falle 1986]{dru86} Drury, L.O'C. \& Falle,
S. A. E. G. 1986, \mnras, 223, 353

\bibitem[Drury, Markewiecz \& V\"olk 1989]{drumv89} 
Drury, L. O'C., Markiewicz, W. J. \& V\"olk, H. J. 1989, \aap, 225, 179

\bibitem[Drury \& V\"olk 1981]{dru81} Drury, L.O'C. \& V\"olk,
H. J. 1981, \apj, 248, 344 

\bibitem[Ellison, M\"obius \& Paschmann 1990]{ell90} Ellison, D. C., M\"obius, E. \& Paschmann, G. 1990, \apj, 352, 376

\bibitem[Frank \etal 1995]{fran95} Frank, A., Jones, T. W. \& Ryu, D. 1995, \apj, 441, 629

\bibitem[Jones \& Ellison 1991]{jonel91} Jones, F. C. \& Ellison, D. C. 1991, Space Sci. Rev., 58, 259

\bibitem[Jones \& Kang 1990]{jon90} Jones, T. W. \& Kang, H. 1990,
\apj, 363, 499 

\bibitem[Jones \& Kang 1992]{jon92} Jones, T. W. \& Kang, H. 1992,
\apj, 396, 575

\bibitem[Jones \& Kang 1993]{jon93} Jones, T. W. \& Kang, H. 1993,
\apj, 402, 560

\bibitem[Jones et al., 1981]{jss81} Jones, E. M., Smith, B. W., \&
Straka, W. C. 1981, \apj, 249, 185 

\bibitem[Jun et al., 1994]{jcn94} Jun, B.-I., Clarke, D. A., \&
Norman, M. L. 1994, \apj, 429, 748

\bibitem[Jun et al., 1995]{jns95} Jun, B.-I., Norman, M. L., \&
Stone, J. M. 1995, \apj, 453, 332

\bibitem[Kang \& Jones 1990]{kan90} Kang, H. \& Jones, T. W. 1990, \apj, 353, 149

\bibitem[Kang et al., 1992]{kjr92} Kang, H., Jones, T. W., \& Ryu, D.
1992, \apj, 385, 193

\bibitem[Kang \& Jones 1995]{kanj95} Kang, H. \& Jones, T. W. 1995, \apj, 447, 944

\bibitem[Kang \& Jones 1996]{kanj96} Kang, H. \& Jones, T. W. 1996,
\apj, (in press)

\bibitem[Krymsky 1977]{kry77} Krymsky, G. F. 1977,
Dok. Acad. Nauk. USSR, 234, 1306 

\bibitem[Malkov \& V\"olk 1996]{malv96}
Malkov, M. A. \& V\"olk H. J. 1996, \apj, (in press)

\bibitem[Ryu, Kang \& Jones 1993]{rkj93}
Ryu, D., Kang, H. \& Jones, T. W. 1993, \apj, 405, 199

\bibitem[Webb 1983]{web83} Webb, G. M. 1983, \aap, 127, 97

\bibitem[Webb \etal 1986]{webb86} Webb, G. M., Drury, L. O'C., \& V\"olk, H. J. 1986, \aap, 160, 335


\end{thebibliography}
\end{document}